\documentclass[10pt]{article}

\usepackage{graphics,subfigure}
\usepackage{amsfonts,amsmath,amssymb}

\newcommand{\tabletext}{\scriptsize}

\newcommand{\Mc}{\mathcal{M}}
\newtheorem{theorem}{Theorem}
\newtheorem{corollary}{Corollary}
\newtheorem{remark}{Remark}
\newcommand{\Pp}{\mathsf{P}}
\newcommand{\R}{\mathbb{R}}
\newcommand{\Tb}{\mathbb{T}}
\newcommand{\N}{\mathbb{N}}

\begin{document}

\title{Large Deviations for Random Trees and the Branching of RNA Secondary
Structures}
\author{Yuri Bakhtin\thanks{Partially supported by NSF CAREER DMS-0742424}\
\and Christine E. Heitsch\thanks{Partially supported by a BWF CASI and NIH NIGMS 1R01GM083621-01}}
\date{Feburary 11, 2008}
\maketitle

\begin{abstract}
We give a Large Deviation Principle (LDP) with explicit rate function 
for the distribution of vertex degrees in plane trees, a combinatorial
model of RNA secondary structures.
We calculate the typical degree distributions based on nearest neighbor
free energies, and compare our 
results with the branching configurations found in two sets of 
large RNA secondary structures. 
We find substantial agreement overall, with some interesting deviations
which merit further study.
\end{abstract}

\section{Introduction}

In this paper we give a Large Deviation Principle (LDP) for a 
combinatorial model of RNA secondary structures.  
This mathematical result allows us to make quantitative statements about
the expected or ``typical'' branching configurations for our model 
of RNA folding.
We are motivated by the question of identifying ``unusual'' substructures
in large RNA molecules, which is a crucial aspect of searching for 
putative functional motifs.
This is a challenging biological question, particularly for lengthy RNA 
sequences whose size is problematic for most existing computational approaches.
We address one aspect of this problem by investigating the asymptotic 
branching degrees of large random trees under distributions which reflect 
the thermodynamics of RNA base pairing. 

Previous combinatorial results~\cite{insights} on plane trees suggest
that the degree of loop branching is correlated with thermodynamic
stability and functional significance.
We refine this analysis of the branching degree in RNA secondary structures
by considering Gibbs distributions based on the nearest neighbor free
energy parameters. 
We are particularly interested in the interplay between the energy
term, which has dominated previous analyses, and the impact of
entropy considerations in determining ``unusual'' configurations.
Our mathematical results are given as an LDP 
for the distribution of vertex degrees among 
plane trees with $N$ vertices. To the best of our knowledge, no studies of Gibbs distributions on random
trees have been published, and our analysis of the energy-entropy competition for these random trees model appears to be new. 
We also compare our expected configurations as $N \rightarrow \infty$ 
with the branching degrees found in two sets of RNA secondary structures:
large subunit 23S ribosomal structures derived by comparative sequence
analysis from the Gutell Lab at UT Austin
and picornaviral structures predicted by free energy minimization
from the Palmenberg Lab at UW Madison.
We find substantial agreement overall between our asymptotic results 
for large random trees and the branching distributions found in the
RNA secondary structures.
This supports our statistical mechanics approach to developing a 
reasonable and mathematically tractable model of large RNA molecules.
Conversely, deviations from our predictions indicate an aspect of 
RNA folding which is not well covered by the model and which merits
further study.

\section{Overview}

A single-stranded RNA sequence encodes molecular structure and function
in a hierarchical way~\cite{tinoco-bustamante-99}, from primary sequence
through secondary structure\footnote{There is a large body of literature on
protein secondary structures (amino acid alpha helices and beta sheets).
However, this is unrelated to the nucleotide base-pairing pattern that
constitutes an RNA secondary structure.}
to the tertiary interactions that determine the three-dimensional structure.
Since the primary structure of an RNA molecule is a nucleotide sequence
much like DNA, experimental sequencing techniques can easily determine 
its base composition, and there are ever-increasing numbers of known 
RNA sequences.
RNA molecules also resemble proteins though, since unlike the canonical 
DNA double helix, different RNA sequences fold into a variety of 
three-dimensional structures. 
However, there are still only a few hundred solved RNA structures,
largely small molecules or molecular fragments, in contrast to the 
thousands of known protein structures.
Thus, understanding the relationship between an RNA sequence and the base
pairings of its secondary structure is an essential step in understanding
the RNA structure-function hierarchy.
Beyond the computational problem of RNA secondary structure determination,
there is the question of evaluating the significance of the base pairings.
In particular, identifying ``unusual'' substructures in large RNA
molecules is a crucial aspect of searching for putative functional motifs.

We begin addressing this problem by investigating the typical branching 
configurations of large RNA molecules using a statistical mechanics approach
with a combinatorial model of RNA folding.
As detailed in~\cite{insights, plane}, trees are widely used to represent
nested RNA secondary structures, and as described in Section~\ref{trees}
we model the folding of RNA sequences using plane trees -- 
ordered, rooted trees~\cite{stanley-99} which nicely
abstract the different substructures in RNA folding.
In Section~\ref{sec:LDP} we consider the set of all plane trees on $N$ vertices and define a 
Gibbs distribution on that set using energy functions from the
nearest neighbor free energy model for RNA folding.
We analyze these distributions as $N\to\infty$,  and give an LDP 
with explicit rate function.

Informally, an LDP with nonnegative rate function $I$
for random variables $X_{N}$ taking values in a set $\Mc$ means that
for all $p \in \mathcal{M}$ and large $N$, we have 
\[ \Pp\{X_{N} \approx p\} \approx e^{-N I(p)} \mbox{.}\]
In particular, when the minimal value $0$ is attained by $I$ at a unique
point $p^{\ast} \in \mathcal{M}$, then for any neighborhood $O$ of 
$p^{\ast}$, the probability $\Pp\{X_N\notin O\}$ decays exponentially in $N$.
This can also be restated as a Law of Large Numbers with exponential 
convergence in probability to the limit point $p^{\ast}$.

As a consequence of this Law of Large Numbers,
it makes sense to call a random tree from our model ``typical'' 
if the distribution of its branching degrees is close to $p^*$. 
More precisely, the LDP for our model tells us that 
there is a distribution $p^*$ of branching degrees such that the 
distribution for a random tree is close to $p^*$
with probability approaching 1 as the size of the tree grows to infinity.
Therefore, it also makes sense
to consider any tree with a branching degree distribution considerably deviating from $p^*$ to be exotic.
In Section~\ref{applications} we compute $p^*$,
the asymptotically most probable branching sequences for our model. 
An immediate implication is that 
it is unlikely (in the framework of our model) to observe a large RNA secondary structure with branching degree distribution that significantly differs from $p^*$. 
However, if such conformation is observed, the analysis of that conformation should result in some new insights.

Under the nearest neighbor thermodynamic model for RNA folding, 
the free energy of an RNA secondary structure is assumed to be the 
independent sum of the substructure free energies.
In our model of RNA branching configurations, this corresponds to an 
assumption that the free energy of the entire tree is equal to the 
sum of free energies associated with each vertex. 
However, it is known from statistical mechanics that free energy is additive if all the subconfigurations are statistically
independent of each other. If this requirement is not satisfied then additional entropy corrections related to the
interdependencies or interactions between the subsystems or subconfigurations should appear.

We show that this is indeed the case for the systems that we consider. The combinatorial structure of the trees imposes certain restrictions
on branching degrees that lead to their mutual statistical dependence which in turn induces certain entropy corrections.  
Due to this interplay between energy and entropy, the 
typical trees minimize the free energy corrected by the extra entropy term resulting from the
combinatorics of plane trees, and do not minimize the energy plainly understood as sum of the energies of all the vertices.
Therefore, the entropy correction is an important factor in determining the branching
of typical large trees, which have a broader distribution of loop degrees
than the exotic energy-minimizing configurations.

Based on our results, we have that the percentage of high degree vertices 
in a typical large tree is exponentially decreasing, but positive. 
As we discuss, the exact rate of decay depends on the specific thermodynamic
parameters, and there are interesting differences in the behavior of our 
model under the two sets of energy values considered. 
In Section~\ref{compres},
we compare these asymptotic degree distributions with the branching found
in a set of ribosomal and a set of picornaviral RNA secondary structures.
There are definite qualitative similarities between our predictions
and the secondary structure data, as well as various differences which 
suggest areas for future investigations.

\begin{figure}
\begin{center}
\includegraphics{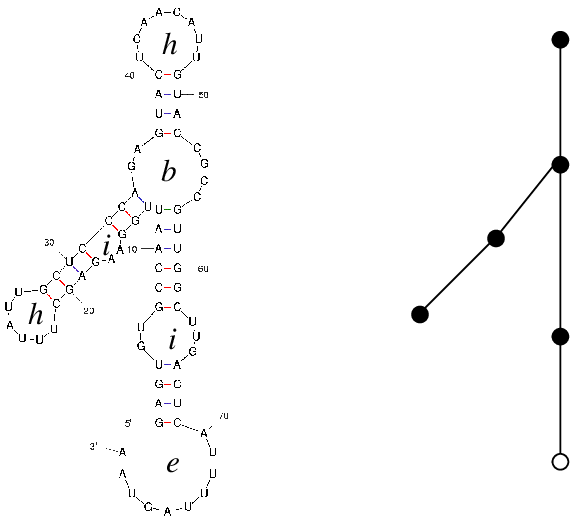}
\end{center}
\caption{The secondary structure, generated by the \texttt{mfold} Web Server
available through \texttt{http://frontend.bioinfo.rpi.edu/zukerm/home.html},
for a 79 base fragment from the 3' UTR of the 7440 nucleotide RNA virus 
poliovirus 1-Mahoney, Genbank Accession No. J0228140~\cite{palmenberg-sgro-97}.
The structure has two hairpin loops, two internal loops (one of which is 
a bulge loop of size 2), one branching (multi) loop, and an external loop.
The adjacent plane tree (rooted at the bottom) models the configuration 
of the RNA secondary structure, preserving information about the basic 
arrangement of loops/vertices and helices/edges.\label{frag}}
\end{figure}

\section{Modeling RNA folding by trees}\label{trees}

As pictured in Figure~\ref{frag},
RNA secondary structures can be modeled as trees by collapsing 
each single-stranded loop into a point and replacing the stacked base 
pairs by an edge connecting two such points.
The tree is rooted at the vertex corresponding to the external loop, 
which contains the 5' and 3' ends of the sequence, and by imposing a 
linear ordering on the vertices of the tree, we maintain the 5' to 3' 
orientation of the RNA molecule.
Such an ordered, rooted tree, known as a plane tree~\cite{stanley-99},
gives a ``low-resolution'' model of RNA folding;
it preserves information about the basic arrangement of loops and helices
in an RNA secondary structure, and also captures certain essential 
elements of the free energy thermodynamic model.
The free energy of a particular RNA secondary structure is calculated
as the independent sum over the energies of well-defined 
substructures~\cite{zuker-mathews-turner-99},
namely the helices and different classes of loop structures.
The primary loop classification is according to the number of base
pairs, that is according to the branching degree (the number of children)
of the corresponding vertex. 
Since we consider only the branching degree of the vertices in our rooted
trees, we will frequently refer to the number of children simply as
the degree of the vertex. 
Hence, there are three basic types of loops which we consider 
with the associated free energies given in Table~\ref{loopdg}.
For our purposes, we consider bulge loops to be a special type of 
internal/degree~1 loop, loops with degree~$\geq 2$ are called ``branching''
loops rather than ``multiloops'', and the exceptional energy function for 
the external loop is disregarded. 

\begin{table} 
\centering
\begin{tabular}{|c|c|c|c|} \hline
Name & Branching degree & $d G$ 2.3 & $d G$ 3.0 \\ \hline
Hairpin & 0 & 3.5 & 4.10 \\ \hline
Internal & 1 & 3.0 & 2.3  \\ \hline
Branching & $d \geq 2$ & 4.6 - 0.2 (d + 1) & 3.4 - 1.5 (d + 1)  \\ \hline
\end{tabular} 
\caption{Loop structures and associated free energies 
at $37^{\circ}$~\cite{mathews-etal-99, zuker-03}.\label{loopdg}}
\end{table}

Here, we consider two possible energy values for each type of loop structure, 
corresponding to the current standard known as $d G$ 3.0 and the former
standard $d G$ 2.3.
(See 
``Version 3.0 free energy parameters for RNA folding at $37^{\circ}$''
and ``Version 2.3 free energy parameters for RNA folding at $37^{\circ}$''
available through the \texttt{mfold} website.) 
The energy of a loop is a function of the number of single-stranded bases
and the number of base pairs, with an additional dependency for the 
stacking interactions~\cite{zuker-mathews-turner-99}.
For the purposes of our model, we have chosen a specific energy value
from the unbounded set of possibilities for each of the three types of
loops.
These values correspond to loops where any enclosed base pairs are G -- C,
the closing base pair is C -- G, and the single-stranded segments are $A^{4}$.
These loops occur in the combinatorial model of RNA folding
previously considered in~\cite{insights},
and the $d G$ 3.0 thermodynamic values were used in the results on RNA 
branching degrees given there.
Clearly, there are many other possible choices and it may be interesting
to investigate the impact of different thermodynamic values -- energy
minimizing versus maximizing, average against frequent, etc. -- on the
behavior of the model.
We note that the $d G$ 2.3 parameters were originally included in our analysis 
because the picornaviral secondary structures from~\cite{palmenberg-sgro-97} 
which are analyzed in Section~\ref{compres} were determined using those values.
In doing so, though, we noticed interesting changes in the evolution of
the free energy model.

The free energy model is evolving in two significant ways.
One type of development is extending and refining the experimental 
determination of thermodynamic values for the entropy and enthalpy of specific
base interactions~\cite{lu-turner-mathews-06, mathews-etal-99}.
While this has improved the accuracy of RNA secondary structure prediction,
it has also greatly increased the complexity of the thermodynamic calculations;
the free energy model now includes more than 10,000 parameters,
nearly all of which pertain to small internal loops.
The other evolving component is changes in the estimation of free energy
functions which have not, or worse cannot, be measured directly.
The loop destabilizing energies are the most notable instance of this,
and the major source of change between the previous energy parameters 
($d G$ 2.3) and those currently used ($d G$ 3.0). 
Through our mathematical results given in the next two sections, though,
we can assess the impact of these changes and the importance of the 
entropy correction on the likely configurations
of large RNA secondary structures without getting lost in the
thousands of detailed thermodynamic parameters.

\section{The Large Deviation Principle}\label{sec:LDP}

As described above, we consider plane trees as our combinatorial model 
of RNA folding.
Now, we introduce a family of Gibbs distributions on the trees, and 
state our main mathematical results.

We fix a number $D\in\N$ and for each $N\in\N$ consider the set $\Tb_N(D)$ of plane trees on $N\in\N$ vertices such that the
number of children of each vertex  (the branching degree) does not exceed $D$.  We restrict ourselves to the trees with
bounded degrees to simplify the mathematical treatment. 
However, if $D$ is suitably large, this does not impose any significant
restrictions since, although the degree of branching in RNA loops is 
theoretically unbounded, in practice it is necessarily limited by 
physical constraints.
Moreover, as we shall see in the next section, the properties of the 
model stabilize as $D\to\infty$.

To define Gibbs distributions on $\Tb_N(D)$ we associate an energy with each plane tree.
In our model of RNA branching configurations,
we assume that the energy associated with each vertex depends only on its  branching degree and is given by
a function $c:\{0,1,\ldots,D\}\to\R$. 
To a first approximation, this is consistent with the thermodynamics of
RNA folding.
The energy of a tree $T\in\Tb_N(D)$ is then given by
\begin{equation}
H(T)=\sum_{j=1}^N c(d_j(T))=\sum_{k=0}^D c(k) \chi_k(T),
\end{equation}
where $d_j$ denotes the branching degree of vertex $j$, and $\chi_k(T)$ is the number of vertices with $k$ children in $T$. 
Now the Gibbs probability measure on $\Tb_N(D)$ associated with $H$ is given by
\begin{equation*}
P_N\{T\}=\frac{e^{-\beta H(T)}}{Z_N},\quad T\in \Tb_N(D),
\end{equation*}
where $\beta>0$ is the inverse temperature parameter  and 
$Z_N$ is a normalizing constant known as the partition function:
\begin{equation*}
Z_N=\sum_{T\in \Tb_N(D)} e^{-\beta H(T)}.
\end{equation*}

There are several interesting questions one could ask about the asymptotic behavior of measures $P_N$ as $N\to\infty$. 
Here we would like to study the frequencies of branching degrees, so 
for each $N$ we
introduce a probability measure $\nu_N$ on $[0,1]^{D+1}$ defined as the distribution
of the random vector $\frac1N (\chi_0(T),\chi_1(T),\ldots,\chi_D(T))$ under $P_N$. Our main result is an LDP
for $\nu_N$.

\medskip

Let us recall that a sequence of probability measures $(\mu_N)_{N\in\N}$ on a compact metric space $(E,\rho)$ satisfies an LDP with a nonnegative lower-semicontinuous 
rate function $I:E\to\R$ if  
\begin{equation*}
\limsup_{N\to\infty}\frac{1}{N} \ln\mu_N(C)\le -I(C),\quad \mbox{\rm for any closed set\ }C\subset E,
\end{equation*}
and
\begin{equation*}
\liminf_{N\to\infty}\frac{1}{N} \ln\mu_N(O)\ge -I(O),\quad \mbox{\rm for any open set\ }O\subset E,
\end{equation*}
where for a set $O$, we denote $I(O)=\inf_{p\in O} I(p)$, 
see \cite[Section II.3]{Ellis:MR2189669} or \cite[Section 1.2]{Dembo-Zeitouni:MR1619036}.

Informally, an LDP means that if we consider random variables $X_N$ with distribution $\mu_{N}$, then for all $p$ and large $N$
we have $$\Pp\{X_N\approx p\}\approx e^{- NI(p)}.$$ In particular, if the minimal value $0$ is attained by $I$ at a unique point~$p^*$, then for any neighborhood $O$ of $p^*$, $\mu_N(O^c)=\Pp\{X\notin O\}$ decays exponentially in $N$. This can be restated as a Law of Large Numbers with exponential convergence in probability to the limit point $p^*$.  

\medskip

For our model, it is natural to formulate the LDP for $\nu_N$ on the set
\begin{equation*}
\Mc=\left\{p\in[0,1]^{D + 1}:\ \sum_{k=0}^D
p_k=1,\ \sum_{k=0}^D k p_k=1\right\}
\end{equation*}
equipped with Euclidean distance.
Though the random vector $\frac1N (\chi_0,\ldots,\chi_D)$ does not belong to $\Mc$,
it is asymptotically close to $\Mc$:
\begin{equation*}
\sum_{k=0}^D\frac{\chi_k}{N}=1,\quad \sum_{k=0}^Dk\frac{\chi_k}{N}=1-\frac{1}{N}.
\end{equation*}

So instead of formulating an LDP for the sequence of random vectors $\frac1N (\chi_0,\ldots,\chi_D)$,
we shall formulate an LDP for a sequence of random vectors that is close to it and belongs to $\Mc$.

Let us introduce 
$J:\Mc\to\R$ via

\begin{equation*}
J(p)=\beta E(p)-h(p),
\end{equation*}
where
\begin{equation}
\label{eq:entropy}
h(p)=-\sum_{k=0}^D p_k\ln p_k
\end{equation}
is the entropy of the probability vector $p=(p_0,\ldots,p_D)$,
and 
\begin{equation*}
E(p)=\sum_{k=0}^D p_kc(k)
\end{equation*}
is the energy associated with $p\in\Mc$.

The function $J$ is strictly convex, and  attains
its minimum on $\Mc$ at a unique point $p^*$.
Let
\begin{equation}
\label{eq:rate_ordered}
I(p)=J(p)-J(p^*).
\end{equation}

For a measure $Q$ on $[0,1]^{D+1}\times\Mc$ we define  $Q^{(1)}$ and $Q^{(2)}$
as the marginal distributions of $Q$ on $[0,1]^{D+1}$ and $\Mc$ respectively. 

\begin{theorem}\label{th:mainLDP-ordered}
 There is a sequence of probability measures $(Q_N)_{N\in\N}$ defined on
$[0,1]^{D+1}\times\Mc$ with the following properties.
\begin{enumerate}
\item For each $N$, we have $Q_N^{(1)}=\nu_N$.
\item For each $N$, 
\begin{equation*}
Q_N\left\{(x,y)\in[0,1]^{D+1}\times\Mc:\sum_{k=0}^D|x_k-y_k|>\frac{1}{N}\right\}=0.
\end{equation*}
\item The sequence $(Q_N^{(2)})_{N\in\N}$ satisfies LDP on $\Mc$ with the rate function $I$ defined
in \eqref{eq:rate_ordered}.
\end{enumerate}
\end{theorem}

\begin{remark}\rm This theorem says that though the random vector $\chi/N$ does not belong to $\Mc$,
one can find another random vector that is, on the one hand, very close to $\chi/N$ and on the other hand
belongs to $\Mc$ and satisfies the LDP.
\end{remark}

An immediate consequence is the following Law of Large Numbers:
\begin{corollary}\label{cr:LLN2} As $N\to\infty$,
\begin{equation*} 
\left(\frac{\chi_0}{N},\frac{\chi_1}{N},\ldots, \frac{\chi_D}{N}\right)\to p^*
\end{equation*}
in probability.
\end{corollary}

\begin{remark}\rm
The statements above show that with high probability the degree
frequencies are close to $p^*$. Note that in most cases $p^*$ is 
not the minimizer of the energy $E$ on $\Mc$.
\end{remark}

We shall now give a sketch of the proof of Theorem \ref{th:mainLDP-ordered}. The proof is based on the 
fact that trees with equal branching degree sequences have equal energy. Therefore,
\begin{equation}
P_N\left\{\chi(T)=n\right\}= \frac{e^{-\beta E\left(n\right)}C(N,n)}{
Z_N},
\label{eq:projection_of_P_N}
\end{equation}
where $n=(n_0,\ldots,n_D)$ and $C(N,n)$ is the number of plane trees of order $N$ 
with $n_k$ nodes of branching degree $k$:  
$$
C(N,n)=\frac{1}{N}\binom{N}{n_0,\ n_1,\ n_2 \ldots }=\frac{1}{N}\frac{N!}{n_0! n_1! n_2! \ldots }
$$
if $n_1+2n_2+\ldots=N-1$, and $0$ otherwise
(see e.g.~Theorem 5.3.10 in \cite{stanley-99}). One can apply the formula
\begin{align*}
C(N,n)=& \exp\left\{N\left(-\sum_{k=0}^D\frac{n_k}{N}\ln \frac{n_k}{N}+ O\left(\frac{\ln N}{N}\right)\right)\right\},\\
      =& \exp\left\{N h\left(\frac{n}{N}\right) + O(\ln N)\right\}, \mbox{\rm\ as\ } N\to\infty.
\end{align*}
which holds true uniformly in $n$, see e.g.\cite[Lemma I.4.4]{Ellis:MR2189669}.

Plugging this into~\eqref{eq:projection_of_P_N}, we get
\begin{align*}
P_N\left\{\frac{\chi(T)}{N}=\frac{n}{N}\right\}&= \frac{e^{-N\left[\beta E\left(\frac{n}{N}\right)-h\left(\frac{n}{N}\right)\right]+O(\ln N)}}{
Z_N},\\
&=\frac{e^{-NJ\left(\frac{n}{N}\right)+O(\ln N)}}{
Z_N},
\end{align*}
which is the desired asymptotics. In fact, the LDP that we claim is a stronger statement and requires extra work to 
complete this argument rigorously. The complete proof along with other random tree models will appear in detail elsewhere~\cite{yb-ceh-2}.

\section{Applications to RNA secondary structure}\label{applications}
In this section we compute the asymptotically most probable branching sequences for
our model under an additional requirement that the coefficients $c(m)$ are given by
$$
c(m)=\begin{cases}A_1,& m=0,\\ A_2,& m=1,\\ A_3-A_4m,& m\ge2,
\end{cases}
$$ 
for some numbers $A_1,A_2,A_3,A_4$. 
Both the $d G$~2.3 and $d G$~3.0 thermodynamic values in Table~\ref{loopdg}
satisfy this requirement, and
we shall address these models in detail in the end of this section.

For this choice of $c(m)$ we have
$$
\beta E(p)= a_1p_0+a_2p_1+\sum_{m=2}^D(a_3-a_4m)p_m,
$$
where 
$$
a_i=\beta A_i=\frac{A_i}{kT},\quad i=1,2,3,4,
$$
$k=1.99$ Cal/mole$\cdot$K being the Boltzmann constant, and $T$ the temperature.

Corollary \ref{cr:LLN2} implies that a typical conformation will have degree frequencies close to the
solution of
\begin{equation*}
S(p)\to\min,\quad
p\in\Mc,
\end{equation*}
where
$$
S(p)=\sum_{m=0}^Dp_m\ln p_m +a_1p_0+a_2p_1+\sum_{m=2}^D(a_3-a_4m)p_m.
$$
It is easy to see that, since the function $x\mapsto x\ln x$ has infinite negative derivative at zero, the minimal value of $S(p)$ cannot be attained at the boundary of $\Mc$. Moreover, $S$ is strictly
convex, so that there is a unique minimizer. Therefore we can solve this problem by the method of
Lagrange multipliers. We set
$$
S(p,\lambda)= S(p)+\lambda_0\left(\sum_{m=0}^Dp_m-1\right)+\lambda_1\left(\sum_{m=0}^Dmp_m-1\right).
$$
The optimal vector $( p^*,\lambda)$ must satisfy
$$
0=\frac{\partial}{\partial p_m}S( p^*,\lambda)=
\begin{cases}
a_1+\ln p_0^* +1 +\lambda_0,&m=0,\\
a_2+\ln p_1^* +1 +\lambda_0+\lambda_1,&m=1,\\
a_3-a_4m+\ln  p_m^*+1 +\lambda_0 +m\lambda_1,&m\ge2.
\end{cases}
$$

We rewrite this as
\begin{equation}\label{eq:system_on_p_mu_nu}
\begin{cases}
p_0^*=b_1^{-1}\mu,\\
p_1^*=b_2^{-1}\mu\nu,\\
p_m^*=b_3^{-1}\mu(b_4\nu)^m,&m\ge2,
\end{cases}
\end{equation}
where $\mu=e^{-\lambda_0-1},\nu=e^{-\lambda_1}$ and $b_i=e^{a_i}, i=1,\ldots,4$.
We notice that
\begin{align*}
1=\sum_{m=0}^Dp_m^*=\mu\left(b_1^{-1}+b_2^{-1}\nu+b_3^{-1}\sum_{m=2}^D(b_4\nu)^m\right),\\
1=\sum_{m=0}^Dmp_m^*=\mu\left(b_2^{-1}\nu+b_3^{-1}\sum_{m=2}^Dm(b_4\nu)^m\right).
\end{align*}
Instead of solving this system explicitly, let us consider the case of $D\gg 1$, i.e., rewrite
the limiting system for $D\to\infty$:
\begin{align*}
1=\mu\left(b_1^{-1}+b_2^{-1}\nu+b_3^{-1}\frac{b_4^2\nu^2}{1-b_4\nu}\right),\\
1=\mu\left(b_2^{-1}\nu+b_3^{-1}\frac{2b_4^2\nu^2-b_4^{3}\nu^3}{(1-b_4\nu)^2}\right).
\end{align*}

Excluding $\mu$ we get a quadratic equation on $\nu$ and among the two roots we choose 
$$\nu=\frac{\sqrt{b_3}}{b_4(\sqrt{b_1}+\sqrt{b_3})}$$
that satisfies
$0<b_4\nu<1$.

Now we can express $\mu$ as
$$\mu={\frac { \left( -b_{{1}}+\sqrt {b_{{3}}b_{{1}}} \right) b_{{4}} \left( -b_{{3}}+b_{{1}} \right)
b_{{2}}b_{{1}}}{-b_{{2}}b_{{4}}b_{{3}}\sqrt {b_{{3}}b_{{1}}}-2\,{b_{{1}}}^{2}b_{{2}}b_{{4}}\\
\mbox{}+3\,b_{{2}}b_{{4}}b_{{1}}\sqrt {b_{{3}}b_{{1}}}+2\,b_{{3}}{b_{{1}}}^{2}-b_{{3}}b_{{1}}\sqrt {b_{{3}}b_{{1}}}\\
\mbox{}-{b_{{1}}}^{2}\sqrt {b_{{3}}b_{{1}}}}}.$$

\bigskip

\begin{figure}
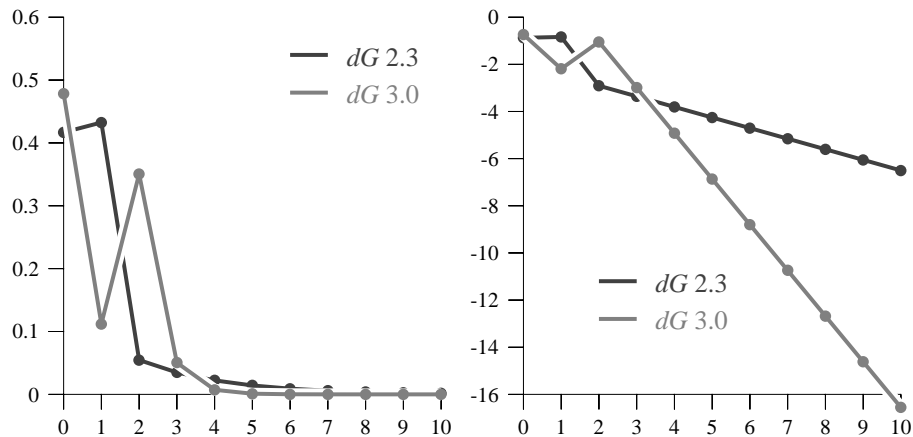

\includegraphics{gexp}
\includegraphics{gexp2}
\caption{The first 11 values of $p_{m}$ for both the $d G$~3.0 model
and the $d G$~2.3 model, where the right-hand graph shows the 
logarithm of the values. \label{gexp}}
\end{figure}

For the $d G$~3.0 model, we have
$A_1= 4.1$ KCal/mole, $A_2=2.3$ KCal/mole, $A_3=1.9$ KCal/mole,
$A_4=1.5$ KCal/mole at $T=273+37=310$ K. Then the solution given above, yields
$$\nu\approx 0.013,\quad\mu\approx 368.3.$$
Likewise, for the $d G$~2.3 model, we have
$A_1= 3.5$ KCal/mole, $A_2=3.0$ KCal/mole, $A_3=4.4$ KCal/mole,
$A_4=0.2$ KCal/mole at $T=273+37=310$ K. Then the solution given above, yields
$$\nu\approx 0.46,\quad\mu\approx 121.3.$$
The first several values of $p_m$ in both cases are displayed in 
Figure~\ref{gexp}.

The LDP for our model implies that, typically,
the frequency of the loops of degree $k$ decreases exponentially in $k$.
However, the relative frequency for the first three terms and the exact 
rate of decay depends on the specific thermodynamic parameters. 
We know from previous results~\cite{insights} that the trees which 
minimize the associated free energies in the $d G$~3.0 model maximize
the number of vertices of degree 2.
We see a similar behavior in the asymptotic distribution of vertex degrees
under our LDP with the $d G$~3.0 thermodynamic values; in a typical large tree,
$47.8\%$ of the vertices would have degree 0 and $35.1\%$ would have degree 2.
Because of the impact of the entropy term correction, though, 
$11.2\%$ of the vertices would have degree 1, and a vanishingly small but 
still nonzero percentage would be likely to have some degree $\geq 3$.
Thus, under the $d G$~3.0 model, the frequency of branching degrees in 
a typical large tree is a refined, and certainly more reasonable, distribution
which still resembles our original calculation of the energy-minimizing 
configurations. 

In contrast, the relative frequency among the vertices with degree 0, 
degree 1, and degree $\geq 2$ is significantly different for the 
distribution calculated with the $d G$~2.3 values.
Now, in a typical large tree, while $41.7\%$ of the vertices would still
have degree 0, only $5.5\%$ would have degree 2, and $43.2\%$ would have
degree 1.
Furthermore, although the percentage of loops with degree $\geq 3$ still
decreases exponentially, the rate is significantly lower than it was 
with the $d G$~3.0 values. 
The differences in the thermodynamic values are primarily a result of
changes in the loop destabilizing energies for the hairpin and internal
loops as well as more significant changes in the offset,
free base penalty, and helix penalty for the multibranched loop energy
function.
In particular, the $d G$~2.3 values for the offset, free base penalty, 
and helix penalty are $4.60$, $0.40$, and $0.10$ respectively, 
while the $d G$~3.0 values are $3.40$, $0.0$, $.40$. 
Intuitively, branching is significantly more favorable, energetically
speaking, under the $d G$~3.0 thermodynamic model than it was in the 
$d G$~2.3 version.
These changes then have a significant impact on the distribution
among loops of small degrees as well as on the decay rate for the tail of
the distribution.

In our model, we are able to assess the impact of these changes 
on the distribution of branching degrees for a typical large tree.
However, our low-resolution model of RNA folding does not permit
any assessment of the correctness of the two thermodynamic models,
as was done in a recent analysis~\cite{doshi-etal-04}.
As we shall see, though, it is the $d G$~2.3 distribution, and not 
the $d G$~3.0 model, which more closely resembles the frequency of 
branching degrees in both the large subunit 23S ribosomal and the 
picornaviral RNA secondary structures.

\section{Ribosomal and picornaviral branching degrees}\label{compres}

We analyze the branching degrees found in two different sets of RNA secondary 
structures, and compare them with the typical branching sequences for
our large random trees.  
Our findings are summarized here in Figure~\ref{dist} and in the discussion,
while more details are given in Appendix~\ref{data}.
Overall, the branching of these secondary structures agrees with the 
results for our model, although there are deviations which suggest 
interesting avenues for further investigation. 
Our comparisons are qualitative, rather than quantitative, since it would
be unrealistic to expect precise agreement between our 
``low-resolution'' model of RNA folding and the branching configurations
of large ribosomal and picornaviral secondary structures.
Still, we find some striking similarities between the predictions based
on our model and the data for real RNA sequences.

\begin{figure}
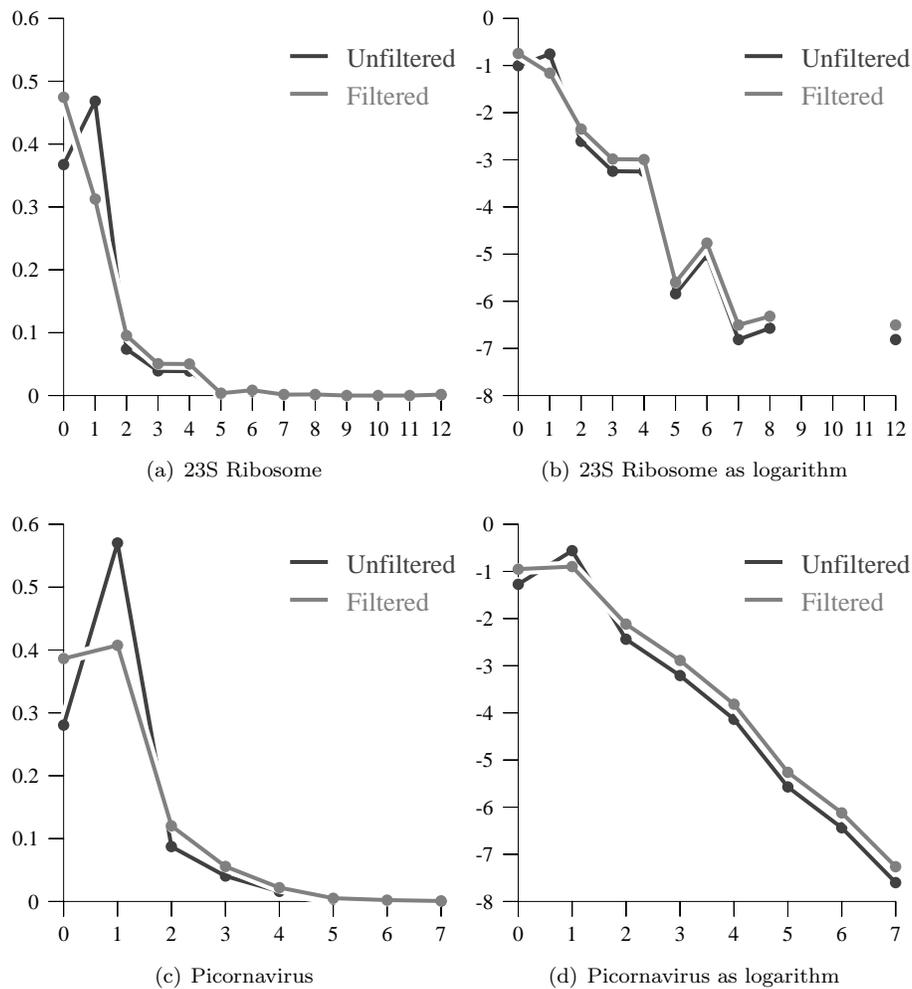

\subfigure[23S Ribosome]{\includegraphics{gribo}}
\subfigure[23S Ribosome as logarithm]{\includegraphics{gribo2}} \\
\subfigure[Picornavirus]{\includegraphics{gpico}}
\subfigure[Picornavirus as logarithm]{\includegraphics{gpico2}}
\caption{The distribution of loop degrees as fractions of the total.  
Each graph shows both the averages over the data set, as given in 
Tables~\ref{l23s_deg_perc} and~\ref{pico_deg_perc},
and the filtered averages, as given in Tables~\ref{l23s_less3_perc}
and~\ref{pico_less3_perc}, after the smallest internal loops have been 
removed. \label{dist}}
\end{figure}

The first set of results, found in Appendix~\ref{L23S}, is for the large 
subunit 23S ribosomal RNA secondary structures determined through 
comparative sequence analysis by the Gutell Lab.
We give results for 20 of the 77 pseudoknot-free sequences available online 
through their Comparative RNA Web (CRW) Site and Project~\cite{cannone-etal-02}.
The chosen sequences were also used in the analyses 
of~\cite{fields-gutell-96} and are representative of the whole set.
As seen in Table~\ref{l23s_seqinfo}, the average sequence length is 
2756.2 nucleotides, although there is certainly variability among the different 
types of ribosomal sequences.
Since our results are asymptotic, we disregard the particular energy function
for the external loop, and the degrees of the external loops are listed
separately in Table~\ref{l23s_seqinfo}.

In Tables~\ref{l23s_degree} and~\ref{l23s_deg_perc}, we give the distribution
of loop degrees, where the degree of a loop is one less than the number
of base pairs contained in the loop.
We see that the most prevalent loops (46.81\% overall) are the internal 
loops with degree 1, followed by the hairpin structures with degree 0. 
Most of the branching loops have degree 2, which agrees with the previous
combinatorial analysis~\cite{insights}, although there is a distribution
extending out to branching loops of degree 12.
We note that the distribution of branching loops tails off much as we 
expected, although there is an interesting peak of degree 6 loops 
as well as smaller peaks at 4, 8, and of course 12.
We find this correlation between loop parity and frequency interesting,
although since ribosomal structure is highly conserved across various 
organisms, the distribution of loop degrees for these 23S RNA secondary 
structures are by no means independent.

As we do for the picornaviral sequences, discussed below, we investigate
in more detail the distribution of sizes among the internal loops.
As we see from Table~\ref{l23s_int}, with only a
few exceptions, the internal loops contain fewer than 16 unpaired bases, 
and a substantial fraction (48.36\% on average) contain at most 2.
It is reasonable~\cite{zuker-jaeger-turner-91} to consider two helices
which are interrupted by an internal loop of fewer than 3 bases as one
contiguous stem.
When we adjust the count of loop degrees accordingly, by excluding 
internal/degree~1 loops with at most 2 unpaired bases as in 
Tables~\ref{l23s_less3} and~\ref{l23s_less3_perc}, then we see
a distribution with different relative numbers of hairpin/degree~0,
internal/degree~1, and branching/degree~$\geq 2$ loops.
For these 23S ribosomal secondary structures, our prediction branching
distributions for $d G$~2.3 are closer to the original unfiltered 
distribution, although the opposite will be true for the picornaviral 
secondary structures. 

The second set of results is found in Appendix~\ref{pico}.
We consider the 11 picornaviral sequences analyzed in~\cite{palmenberg-sgro-97},
which are available online from the Palmenberg Lab 
through \texttt{http://www.virology.wisc.edu/acp/RNAFolds}.
The predicted secondary structures were computed by the \texttt{mfold} 
program v2.2, using the default values~\cite{palmenberg-sgro-97}.
The average length for these sequences, as seen from Table~\ref{pico_seqinfo},
is 7566.27 bases -- considerably longer than the large subunit 23S ribosomal 
sequences.
We also list the external loop degrees separately in Table~\ref{pico_seqinfo},
since this special energy function is not considered in our asymptotic 
results.

Again, the most prevalent loops have degree 1, as seen in 
Tables~\ref{pico_degree} and~\ref{pico_deg_perc}, and the most common type
of internal loops (48.14\% on average) are those containing at most 2 
unpaired bases.
However, the relative number of hairpin/degree~0, internal/degree~1,
and branching/degree~$\geq 2$ loops given in Tables~\ref{pico_degree} 
and~\ref{pico_deg_perc} differs significantly from the LDP distribution.
A large part of this deviation is resolved after further investigation
into the distribution of internal loop sizes.
As seen in Tables~\ref{pico_regint} and~\ref{pico_exint}, there 
is a much broader distribution for the sizes of internal loops.  
While most contain fewer than 16 unpaired bases, the number of ``large''
internal loops does not drop off as sharply for the picornaviral secondary
structures as it did for the 23S ribosomal ones.
When we filter the data by excluding the smallest internal loops, as in 
Tables~\ref{pico_less3} and~\ref{pico_less3_perc}, then we see a 
distribution that agrees even more closely with our LDP probabilities.
In this case, though, we have nearly equal numbers of hairpin/degree~0
loops and internal/degree~1 loops, while the numbers of 
branching/degree~$\geq 2$ loops drop off almost by a factor of 2.
Thus, the predicted picornaviral configurations are less extensively
branched than the ribosomal secondary structures, and the degree of 
branching more closely agrees with our LDP probabilities for the 
$d G$~2.3 model.

\section{Discussion of related results}\label{discuss}

We adopt here a statistical mechanics approach, not to predict base pairs
for a particular RNA sequence, but to analyze what a typical branching 
distribution might be for an arbitrary large RNA secondary structure.
This work joins a growing body of results which analyze different 
general characteristics of RNA secondary structures, both 
theoretically~\cite{clote-etal-06, hofacker-schuster-stadler-98, nebel-04a, nebel-04b} and 
computationally~\cite{clote-etal-05a, fontana-etal-93b, miklos-meyer-nagy-04, rivas-eddy-00, workman-krogh-99, wuchty-etal-99}.
The qualities investigated have been 
the free energy and molecular stability~\cite{clote-etal-05a, miklos-meyer-nagy-04, rivas-eddy-00, workman-krogh-99, wuchty-etal-99} 
as well as the number and type of different 
substructural elements~\cite{clote-etal-06, fontana-etal-93b, 
hofacker-schuster-stadler-98, nebel-04a, nebel-04b}. 
Asymptotics of the expected maximum number of base pairs are studied 
in~\cite{clote-etal-06}, but the overall molecular configurations are
not addressed.

Statistics for different structural elements are computed for short 
RNA sequences $\leq 100$ bases in~\cite{fontana-etal-93b}.
The unfiltered distribution of picornaviral degrees agrees closely 
with their statistical reference probability densities, whereas the 
distribution of the 23S ribosomal degrees resembles their ``natural'' 
sequence distribution by having slightly more hairpin~/~degree~0 loops
and fewer internal~/~degree~1 loops.
The statistics of average branching degree given in~\cite{fontana-etal-93b} 
reflect
the fact that for large RNA sequences the size $N$ of the associated tree is, typically, also large. 
Therefore, the average branching degree is close to 1 due to the identity
\[ \sum_{k = 0}^{D} k \frac{\chi_{k}}{N} = 1 - \frac{1}{N}. \]
This also agrees with the
 theoretical limit given in~\cite{hofacker-schuster-stadler-98};
the asymptotic average branching degree of 1 was 
derived for non-root vertices using a model of RNA secondary
structures at the base level and complicated recursion formulae depending
on $n$, the number of bases in the sequence.
We have not yet investigated the other characteristics analyzed 
in~\cite{fontana-etal-93b} and~\cite{hofacker-schuster-stadler-98},
however it may be possible to extend our low-resolution 
model of RNA folding and this statistical mechanics approach to other 
properties of RNA secondary structures.

In~\cite{nebel-04a}, the typical configuration of large subunit ribosomal
RNA is investigated using a approach based on generating functions and
stochastic context-free grammars.
This approach yields explicit formulas for the frequency of different
structural elements as a function of the sequence length $n$.
Using the average sequences lengths for the 23S ribosomal and picornaviral
secondary structures as $n_{1}$ and $n_{2}$, we computed the predicted
number of hairpin, internal, and branching loops as well as the
average degree of a branching loop.
As in~\cite{nebel-04a}, we compare the averages from the RNA
secondary structures and the predicted frequencies, and find reasonably good
agreement for the 23S ribosomal structures.
The relative differences for the predicted frequencies from the unfiltered 23S
ribosomal averages are:
$-3.28\%$ for hairpin loops,
$-13.05\%$ for internal loops,
$1.09\%$ for branching loops,
$-6.35\%$ for the total number of loops,
and $1.84\%$ for the average branching degree.
In contrast, the comparisons for the picornaviral secondary
structures are not as good.
The relative differences for the predicted frequencies from the unfiltered 
picornaviral averages are:
$21.67\%$ for hairpin loops,
$-31.46\%$ for internal loops,
$6.94\%$ for branching loops,
$-10.52\%$ for the total number of loops,
and $19.42\%$ for the average branching degree.
Since the equations in~\cite{nebel-04a} were derived by training
the grammar on a database of large subunit ribosomal RNA,
it is perhaps not surprising that the predictions of the model 
do not correspond as well to the picornaviral secondary structures.
The paper~\cite{nebel-04b} provides related results by considering
a model of RNA folding where two bases pair with probability $p$ and
investigates different properties of the RNA secondary structures,
but not does not include an analysis of branching degrees.

\section{Conclusions}\label{concl}

We considered Gibbs distributions for our plane tree model of RNA folding
based on the nearest neighbor thermodynamics. 
An important feature of our model is that we can describe the typical 
branching configurations of the trees by calculating the asymptotic 
degree sequences via a Large Deviation Principle (LDP). 
As discussed, this has at least two implications for the branching of 
large RNA secondary structures, such as the large subunit 23S ribosomal 
molecules or RNA viral genomes like picornaviruses.

One implication concerns the asymptotic distribution of vertex 
degrees in a large random tree from our model.
The LDP for our model implies that, typically,
the frequency of the loops of degree $k$ decreases exponentially in $k$.
The exact rate of decay depends on the specific thermodynamic parameters,
however, and we considered two sets of energy values, the current standard
$d G$~3.0 and the former standard $d G$~2.3.
Surprisingly, we find that the typical distribution based on the 
$d G$ 2.3 parameters corresponds more closely to the branching degrees
of both the picornaviral and ribosomal RNA secondary structures.
The differences in the thermodynamic values are primarily a result of 
changes in the loop destabilizing energies for the hairpin and internal 
loops as well as more significant changes in the offset,
free base penalty, and helix penalty for the multibranched loop energy 
function.
These changes then have a significant impact on the distribution
among loops of small degrees as well as on the decay rate for the tail of
the distribution.
To be able to distinguish unusual substructures against the background
of a typical configuration, we will need to understand better the 
impact of different thermodynamic values on the behavior of the model. 

A second implication to emerge from our current analysis is 
that combinatorial constraints lead to important
entropy considerations  
in determining the most likely branching
distributions in large random trees.
The nontrivial combinatorics of the plane trees implies that typical trees 
are minimizers of the free energy corrected by an extra entropy term. 
Thus, although the typical trees in the $d G$~3.0 model are structurally,
and therefore energetically, related to the trees which have minimal 
energy, a typical large tree will not be a minimizer 
of the free energy understood as the sum of the energies
of individual loops.
In fact, the LDP tells us that, in our combinatorial model of RNA 
folding, the energy-minimizing trees are extremely improbable.
Thus, when modeling the folding of large RNA molecules, it is important 
to include entropy considerations which distinguish the most likely 
configurations from those which simply minimize the additive free energy.

\section{Acknowledgments}

The authors thank the anonymous reviewers whose comments significantly
improved the paper, and the ABC Math Program in the School of Mathematics 
at Georgia Tech for fostering interdisciplinary research linking 
mathematics with the biological sciences.  

This research of Yuri Bakhtin, Ph.D., is supported in part by 
NSF CAREER DMS-0742424.
This research of Christine E. Heitsch, Ph.D., is supported in part by a 
Career Award at the Scientific Interface (CASI) from the Burroughs 
Wellcome Fund (BWF) and by NIH NIGMS 1R01GM083621-01.

\newpage

\appendix

\section{Analysis of RNA Branching Degrees}\label{data}

\subsection{23S Ribosomal RNA}\label{L23S}

\begin{table}[!ht]
\tabletext
\begin{tabular}{|*{6}{l|}} \hline
Index & Type & Organism Name & GenBank Accession \# & Length & Degree \\ \hline
 1 & a & Haloarcula marismortui & X13738 & 2925 & 1 \\ 
 2 & a & Thermococcus celer & M67497 & 3029 & 1 \\ 
 3 & b & Thermotoga maritima & M67498 & 3023 & 1 \\ 
 4 & b & Thermus thermophilus & X12612 & 2915 & 1 \\ 
 5 & b & Borrelia burgdorferi & M88330 & 2926 & 1 \\ 
 6 & b & Escherichia coli & J01695 & 2904 & 1 \\ 
 7 & b & Pseudomonas aeruginosa & Y00432 & 2893 & 1 \\ 
 8 & b & Bacillus subtilis & K00637 & 2927 & 1 \\ 
   &   & & AF008220 Z99119 & & \\
 9 & b & Mycobacterium leprae & X56657 & 3122 & 1 \\ 
 10 & c & Chlamydomonas reinhardtii & X15727 & 2902 & 1 \\ 
 11 & c & Zea mays & Z00028 & 2985 & 1 \\ 
 12 & m & Chlamydomonas eugametos & AF008237 & 1915 & 13 \\ 
 13 & m & Saccharomyces cerevisiae & J01527 & 3273 & 1 \\ 
 14 & m & Zea mays & K01868 & 3514 & 6 \\ 
 15 & m & Caenorhabditis elegans & X54252 & 953 & 8 \\ 
 16 & m & Drosophila melanogaster & X53506 & 1335 & 9 \\ 
 17 & m & Xenopus laevis & M10217 & 1640 & 12 \\ 
 18 & e & Giardia intestinalis & X52949 & 2850 & 10 \\ 
 19 & e & Saccharomyces cerevisiae & U53879 & 3554 & 7 \\ 
 20 & e & Arabidopsis thaliana & X52320 & 3539 & 12 \\ \hline 
\end{tabular}

\caption{Sequence information, including the degree of the external loop,
 for 20 of the 77 pseudoknot-free 23S ribosomal 
RNA secondary structures from the CRW~~\cite{cannone-etal-02}. 
The 20 selected were also used 
in the analyses of~\cite{fields-gutell-96}, and are representative of the
whole set.  The different types of sequences are (a) Archae, (b) Eubacteria,
(c) Choloroplast, (m) Mitochondria, and (e) Eucarya. \label{l23s_seqinfo}}
\end{table}

\begin{table}[!ht]
\tabletext
\begin{tabular}{|c|c|*{12}{c}c|} \hline
Index & sum & 0 & 1 & 2 & 3 & 4 & 5 & 6 & 7 & 8 & 9 & 10 & 11 & 12\\  \hline
 1 & 197 & 70 & 93 & 14 & 9 & 9 &  & 2 &  &  &  &  &  & \\ 
 2 & 191 & 73 & 88 & 13 & 7 & 6 & 1 & 1 &  & 1 &  &  &  & 1\\ 
 3 & 201 & 72 & 94 & 15 & 9 & 9 &  & 1 & 1 &  &  &  &  & \\ 
 4 & 192 & 72 & 91 & 12 & 7 & 6 & 1 & 1 &  & 1 &  &  &  & 1\\ 
 5 & 201 & 71 & 95 & 15 & 9 & 9 &  & 2 &  &  &  &  &  & \\ 
 6 & 199 & 70 & 95 & 14 & 10 & 8 &  & 1 & 1 &  &  &  &  & \\ 
 7 & 199 & 70 & 95 & 14 & 10 & 8 &  & 1 & 1 &  &  &  &  & \\ 
 8 & 207 & 71 & 102 & 14 & 9 & 9 &  & 1 & 1 &  &  &  &  & \\ 
 9 & 205 & 74 & 100 & 14 & 7 & 6 & 1 & 1 &  & 1 &  &  &  & 1\\ 
 10 & 202 & 70 & 98 & 14 & 9 & 9 &  & 2 &  &  &  &  &  & \\ 
 11 & 206 & 71 & 100 & 15 & 9 & 9 &  & 2 &  &  &  &  &  & \\ 
 12 & 115 & 49 & 50 & 6 & 3 & 5 & 1 & 1 &  &  &  &  &  & \\ 
 13 & 157 & 59 & 73 & 12 & 6 & 3 & 1 & 2 &  &  &  &  &  & 1\\ 
 14 & 180 & 65 & 85 & 15 & 6 & 6 & 1 & 2 &  &  &  &  &  & \\ 
 15 & 49 & 23 & 18 & 4 & 1 & 3 &  &  &  &  &  &  &  & \\ 
 16 & 77 & 33 & 33 & 4 & 1 & 6 &  &  &  &  &  &  &  & \\ 
 17 & 101 & 41 & 46 & 6 & 3 & 3 & 2 &  &  &  &  &  &  & \\ 
 18 & 190 & 74 & 82 & 17 & 7 & 8 & 1 & 1 &  &  &  &  &  & \\ 
 19 & 221 & 80 & 102 & 21 & 8 & 8 &  & 1 &  & 1 &  &  &  & \\ 
 20 & 218 & 80 & 102 & 20 & 7 & 6 & 1 & 1 &  & 1 &  &  &  & \\ \hline
total & 3508 & 1288 & 1642 & 259 & 137 & 136 & 10 & 23 & 4 & 5 & 0 & 0 & 0 & 4\\ \hline 
\end{tabular}

\caption{Degree distributions of loops.\label{l23s_degree}}
\end{table}

\begin{table}[!ht]
\tabletext
\begin{tabular}{|c|*{12}{c}c|} \hline
Index & 0 & 1 & 2 & 3 & 4 & 5 & 6 & 7 & 8 & 9 & 10 & 11 & 12\\  \hline
1 & 35.5 & 47.2 & 7.1 & 4.6 & 4.6 &  & 1.0 &  &  &  &  &  & \\ 
2 & 38.2 & 46.1 & 6.8 & 3.7 & 3.1 & 0.5 & 0.5 &  & 0.5 &  &  &  & 0.5\\ 
3 & 35.8 & 46.8 & 7.5 & 4.5 & 4.5 &  & 0.5 & 0.5 &  &  &  &  & \\ 
4 & 37.5 & 47.4 & 6.2 & 3.6 & 3.1 & 0.5 & 0.5 &  & 0.5 &  &  &  & 0.5\\ 
5 & 35.3 & 47.3 & 7.5 & 4.5 & 4.5 &  & 1.0 &  &  &  &  &  & \\ 
6 & 35.2 & 47.7 & 7.0 & 5.0 & 4.0 &  & 0.5 & 0.5 &  &  &  &  & \\ 
7 & 35.2 & 47.7 & 7.0 & 5.0 & 4.0 &  & 0.5 & 0.5 &  &  &  &  & \\ 
8 & 34.3 & 49.3 & 6.8 & 4.3 & 4.3 &  & 0.5 & 0.5 &  &  &  &  & \\ 
9 & 36.1 & 48.8 & 6.8 & 3.4 & 2.9 & 0.5 & 0.5 &  & 0.5 &  &  &  & 0.5\\ 
10 & 34.7 & 48.5 & 6.9 & 4.5 & 4.5 &  & 1.0 &  &  &  &  &  & \\ 
11 & 34.5 & 48.5 & 7.3 & 4.4 & 4.4 &  & 1.0 &  &  &  &  &  & \\ 
12 & 42.6 & 43.5 & 5.2 & 2.6 & 4.3 & 0.9 & 0.9 &  &  &  &  &  & \\ 
13 & 37.6 & 46.5 & 7.6 & 3.8 & 1.9 & 0.6 & 1.3 &  &  &  &  &  & 0.6\\ 
14 & 36.1 & 47.2 & 8.3 & 3.3 & 3.3 & 0.6 & 1.1 &  &  &  &  &  & \\ 
15 & 46.9 & 36.7 & 8.2 & 2.0 & 6.1 &  &  &  &  &  &  &  & \\ 
16 & 42.9 & 42.9 & 5.2 & 1.3 & 7.8 &  &  &  &  &  &  &  & \\ 
17 & 40.6 & 45.5 & 5.9 & 3.0 & 3.0 & 2.0 &  &  &  &  &  &  & \\ 
18 & 38.9 & 43.2 & 8.9 & 3.7 & 4.2 & 0.5 & 0.5 &  &  &  &  &  & \\ 
19 & 36.2 & 46.2 & 9.5 & 3.6 & 3.6 &  & 0.5 &  & 0.5 &  &  &  & \\ 
20 & 36.7 & 46.8 & 9.2 & 3.2 & 2.8 & 0.5 & 0.5 &  & 0.5 &  &  &  & \\ \hline
total & 36.72 & 46.81& 7.38& 3.91& 3.88& 0.29& 0.66& 0.11& 0.14& 0& 0 & 0 & 0.11\\ \hline
\end{tabular}

\caption{Degree distributions of loops as percentages. \label{l23s_deg_perc}}
\end{table}

\begin{table}[!ht]
\tabletext
\begin{tabular}{|c|*{14}{c}c|c|} \hline
 & \multicolumn{15}{c|}{Number of internal loops with $1 \leq \mbox{ size } \leq 15$} & List of large \\
Index & 1 &  2 &  3 &  4 &  5 &  6 &  7 &  8 &  9 &  10 &  11 &  12 &  13 &  14 &  15 & loop sizes \\  \hline
 1  & 27 & 13 & 6 & 7 & 4 & 7 & 4 & 6 & 10 & 4 & 1 & 1 & 1 & 1 & & 37 \\ 
 2  & 26 & 15 & 2 & 8 & 5 & 6 & 5 & 6 & 9 & 2 & 1 & 1 & & 1 & & 35 \\ 
 3  & 24 & 17 & 4 & 10 & 4 & 10 & 7 & 4 & 6 & 4 & 2 & & & 1 & & 31\\ 
 4  & 27 & 17 & 4 & 11 & 6 & 8 & 4 & 3 & 5 & 2 & 2 & & & 1 & & 31\\ 
 5  & 26 & 18 & 4 & 10 & 7 & 8 & 5 & 4 & 7 & 2 & 1 & 1 & & 1 & & 30 \\ 
 6  & 24 & 19 & 3 & 11 & 8 & 7 & 3 & 5 & 9 & 2 & 2 & & & 1 & & 30 \\ 
 7  & 26 & 17 & 5 & 11 & 6 & 8 & 5 & 4 & 7 & 2 & 2 & & & 1 & & 31 \\ 
 8  & 32 & 18 & 4 & 10 & 7 & 9 & 5 & 4 & 6 & 2 & 2 & & & 1 & 1 & 30 \\ 
 9  & 29 & 19 & 3 & 10 & 8 & 9 & 6 & 3 & 5 & 3 & 2 & & & 1 & 1 & 31\\ 
 10  & 29 & 19 & 3 & 9 & 6 & 10 & 4 & 4 & 7 & 2 & 2 & & & 1 & & 29, 41\\ 
 11  & 28 & 18 & 6 & 10 & 7 & 6 & 7 & 5 & 7 & 1 & 2 & & & 1 & & 20, 29\\ 
 12  & 15 & 9 & 4 & 4 & 1 & 4 & 2 & 1 & 2 & 3 & 1 & & & & 1 & 18, 20, 27\\ 
 13  & 27 & 12 & 4 & 6 & 3 & 5 & & 4 & 4 & & 2 & & 1 & 1 & & 28, 36, 50, 64\\ 
 14  & 26 & 16 & 4 & 9 & 3 & 5 & 5 & 4 & 7 & 2 & 1 & 1 & & 1 & & 31\\ 
 15  & 4 & 8 & 2 & 1 & 1 & & & & 2 & & & & & & & \\ 
 16  & 10 & 9 & 3 & 3 & & 1 & & 1 & 4 & & & & & & & 17, 20 \\ 
 17  & 15 & 9 & 5 & 4 & 2 & 3 & & 1 & 2 & 1 & & & 1 & & & 18, 33, 47\\ 
 18  & 27 & 14 & 4 & 9 & 2 & 5 & 5 & 5 & 3 & 3 & 1 & 1 & & 1 & & 25, 27 \\ 
 19  & 34 & 19 & 3 & 11 & 5 & 7 & 7 & 4 & 5 & 3 & 1 & 1 & & 1 & & 36 \\ 
 20  & 35 & 17 & 4 & 9 & 5 & 7 & 7 & 3 & 5 & 3 & 1 & 1 & & 3 & & 16, 37\\ 
\hline
\end{tabular}

\caption{Number of internal loops of different sizes, given as the distribution
of loops with at most 15 unpaired bases and as a list of large internal 
loop sizes with multiplicity.\label{l23s_int}}
\end{table}

\begin{table}[!ht]
\tabletext
\begin{tabular}{|c|c|*{12}{c}c|} \hline
Index & sum & 0 & 1 & 2 & 3 & 4 & 5 & 6 & 7 & 8 & 9 & 10 & 11 & 12\\  \hline
 1 & 157 & 70 & 53 & 14 & 9 & 9 &  & 2 &  &  &  &  &  & \\ 
 2 & 150 & 73 & 47 & 13 & 7 & 6 & 1 & 1 &  & 1 &  &  &  & 1\\ 
 3 & 160 & 72 & 53 & 15 & 9 & 9 &  & 1 & 1 &  &  &  &  & \\ 
 4 & 148 & 72 & 47 & 12 & 7 & 6 & 1 & 1 &  & 1 &  &  &  & 1\\ 
 5 & 157 & 71 & 51 & 15 & 9 & 9 &  & 2 &  &  &  &  &  & \\ 
 6 & 156 & 70 & 52 & 14 & 10 & 8 &  & 1 & 1 &  &  &  &  & \\ 
 7 & 156 & 70 & 52 & 14 & 10 & 8 &  & 1 & 1 &  &  &  &  & \\ 
 8 & 157 & 71 & 52 & 14 & 9 & 9 &  & 1 & 1 &  &  &  &  & \\ 
 9 & 157 & 74 & 52 & 14 & 7 & 6 & 1 & 1 &  & 1 &  &  &  & 1\\ 
 10 & 154 & 70 & 50 & 14 & 9 & 9 &  & 2 &  &  &  &  &  & \\ 
 11 & 160 & 71 & 54 & 15 & 9 & 9 &  & 2 &  &  &  &  &  & \\ 
 12 & 91 & 49 & 26 & 6 & 3 & 5 & 1 & 1 &  &  &  &  &  & \\ 
 13 & 118 & 59 & 34 & 12 & 6 & 3 & 1 & 2 &  &  &  &  &  & 1\\ 
 14 & 138 & 65 & 43 & 15 & 6 & 6 & 1 & 2 &  &  &  &  &  & \\ 
 15 & 37 & 23 & 6 & 4 & 1 & 3 &  &  &  &  &  &  &  & \\ 
 16 & 58 & 33 & 14 & 4 & 1 & 6 &  &  &  &  &  &  &  & \\ 
 17 & 77 & 41 & 22 & 6 & 3 & 3 & 2 &  &  &  &  &  &  & \\ 
 18 & 149 & 74 & 41 & 17 & 7 & 8 & 1 & 1 &  &  &  &  &  & \\ 
 19 & 168 & 80 & 49 & 21 & 8 & 8 &  & 1 &  & 1 &  &  &  & \\ 
 20 & 166 & 80 & 50 & 20 & 7 & 6 & 1 & 1 &  & 1 &  &  &  & \\ \hline 
total & 2714 & 1288 & 848 & 259 & 137 & 136 & 10 & 23 & 4 & 5 & 0 & 0 & 0 & 4\\  \hline
\end{tabular}

\caption{Degree distributions of loops with contiguous stems. \label{l23s_less3}}
\end{table}

\begin{table}[!ht]
\tabletext
\begin{tabular}{|c|*{12}{c}c|} \hline
Index & 0 & 1 & 2 & 3 & 4 & 5 & 6 & 7 & 8 & 9 & 10 & 11 & 12\\ \hline 
 1 & 39.3 & 29.8 & 7.9 & 5.1 & 5.1 &  & 1.1 &  &  &  &  &  & \\ 
 2 & 41.2 & 26.6 & 7.3 & 4.0 & 3.4 & 0.6 & 0.6 &  & 0.6 &  &  &  & 0.6\\ 
 3 & 40.7 & 29.9 & 8.5 & 5.1 & 5.1 &  & 0.6 & 0.6 &  &  &  &  & \\ 
 4 & 41.4 & 27.0 & 6.9 & 4.0 & 3.4 & 0.6 & 0.6 &  & 0.6 &  &  &  & 0.6\\ 
 5 & 40.8 & 29.3 & 8.6 & 5.2 & 5.2 &  & 1.1 &  &  &  &  &  & \\ 
 6 & 40.0 & 29.7 & 8.0 & 5.7 & 4.6 &  & 0.6 & 0.6 &  &  &  &  & \\ 
 7 & 40.0 & 29.7 & 8.0 & 5.7 & 4.6 &  & 0.6 & 0.6 &  &  &  &  & \\ 
 8 & 42.3 & 31.0 & 8.3 & 5.4 & 5.4 &  & 0.6 & 0.6 &  &  &  &  & \\ 
 9 & 43.5 & 30.6 & 8.2 & 4.1 & 3.5 & 0.6 & 0.6 &  & 0.6 &  &  &  & 0.6\\ 
 10 & 41.2 & 29.4 & 8.2 & 5.3 & 5.3 &  & 1.2 &  &  &  &  &  & \\ 
 11 & 41.3 & 31.4 & 8.7 & 5.2 & 5.2 &  & 1.2 &  &  &  &  &  & \\ 
 12 & 25.3 & 13.4 & 3.1 & 1.5 & 2.6 & 0.5 & 0.5 &  &  &  &  &  & \\ 
 13 & 33.0 & 19.0 & 6.7 & 3.4 & 1.7 & 0.6 & 1.1 &  &  &  &  &  & 0.6\\ 
 14 & 36.9 & 24.4 & 8.5 & 3.4 & 3.4 & 0.6 & 1.1 &  &  &  &  &  & \\ 
 15 & 11.2 & 2.9 & 1.9 & 0.5 & 1.5 &  &  &  &  &  &  &  & \\ 
 16 & 16.6 & 7.0 & 2.0 & 0.5 & 3.0 &  &  &  &  &  &  &  & \\ 
 17 & 21.1 & 11.3 & 3.1 & 1.5 & 1.5 & 1.0 &  &  &  &  &  &  & \\ 
 18 & 41.8 & 23.2 & 9.6 & 4.0 & 4.5 & 0.6 & 0.6 &  &  &  &  &  & \\ 
 19 & 48.5 & 29.7 & 12.7 & 4.8 & 4.8 &  & 0.6 &  & 0.6 &  &  &  & \\ 
 20 & 48.2 & 30.1 & 12.0 & 4.2 & 3.6 & 0.6 & 0.6 &  & 0.6 &  &  &  & \\ \hline
total & 47.46 & 31.25 & 9.54 & 5.05 & 5.01 & 0.37 & 0.85 & 0.15 & 0.18 & 0 & 0 & 0 & 0.15\\ \hline
\end{tabular}

\caption{Degree distributions of loops as percentages with contiguous stems. \label{l23s_less3_perc}}
\end{table}

\clearpage 

\subsection{Picornaviral RNA}\label{pico}

\begin{table}[!ht]
\tabletext
\begin{tabular}{|*{6}{l|}} \hline
Index & Virus Name & GenBank Acc. \# & Length & Degree \\ \hline
 1 & coxsackievirus B3 & M33854 & 7396 & 13 \\  
 2 & ECHO virus-22 & L02971 & 7339 & 17 \\ 
 3 & encephalomyocarditis virus-A & M81861 & 7735 & 25 \\ 
 4 & foot-and-mouth disease virus-A12 & M10975 & 8214 & 1 \\  
 5 & hepatitis A virus-Hml75 & M14707 & 7478 & 20 \\ 
 6 & rhinovirus-14 & K02121 & 7212 & 16 \\ 
 7 & rhinovirus-16 & L24917 & 7124 & 11 \\ 
 8 & Mengovirus-M & L22089 & 7761 & 26 \\ 
 9 & poliovirus 1-Mahoney & J0228140 & 7440 & 20 \\ 
 10 & poliovirus 3-Sabin & X00596 & 7432 & 22 \\ 
 11 & Theiler's murine encephalomyelitis virus-Bean & M16020 & 8098 & 37 \\ \hline 
\end{tabular}

\caption{Sequence information, including the degree of the external loop,
for the 11 picornaviral sequences analyzed 
in~\cite{palmenberg-sgro-97}, available online through 
\texttt{http://www.virology.wisc.edu/acp/RNAFolds}. \label{pico_seqinfo}}
\end{table}

\begin{table}[!ht]
\tabletext
\begin{tabular}{|c|c|*{7}{c}c|} \hline
Index & sum & 0 & 1 & 2 & 3 & 4 & 5 & 6 & 7\\ \hline
 1 & 499 & 142 & 281 & 41 & 20 & 12 & 3 &  & \\ 
 2 & 478 & 130 & 275 & 42 & 23 & 7 & 1 &  & \\ 
 3 & 530 & 138 & 320 & 44 & 16 & 11 & 1 &  & \\ 
 4 & 594 & 167 & 337 & 45 & 25 & 14 & 3 & 1 & 2\\ 
 5 & 482 & 136 & 267 & 58 & 12 & 4 & 3 & 2 & \\ 
 6 & 454 & 126 & 259 & 37 & 25 & 5 & 2 &  & \\ 
 7 & 456 & 140 & 237 & 46 & 23 & 6 & 1 & 3 & \\ 
 8 & 494 & 130 & 300 & 37 & 17 & 8 & 1 & 1 & \\ 
 9 & 485 & 143 & 267 & 43 & 21 & 8 & 2 &  & 1\\ 
 10 & 507 & 137 & 293 & 50 & 20 & 4 & 2 & 1 & \\ 
 11 & 537 & 157 & 309 & 38 & 21 & 9 & 2 & 1 & \\ \hline
total & 5516 & 1546 & 3145 & 481 & 223 & 88 & 21 & 9 & 3\\ \hline
\end{tabular}

\caption{Degree distributions of loops. \label{pico_degree}}
\end{table}

\begin{table}[!ht]
\tabletext
\begin{tabular}{|c|*{7}{c}c|} \hline
Index & 0 & 1 & 2 & 3 & 4 & 5 & 6 & 7\\ \hline
 1 & 26.4 & 52.3 & 7.6 & 3.7 & 2.2 & 0.6 &  & \\ 
 2 & 24.2 & 51.2 & 7.8 & 4.3 & 1.3 & 0.2 &  & \\ 
 3 & 25.7 & 59.6 & 8.2 & 3.0 & 2.0 & 0.2 &  & \\ 
 4 & 31.1 & 62.8 & 8.4 & 4.7 & 2.6 & 0.6 & 0.2 & 0.4\\ 
 5 & 25.3 & 49.7 & 10.8 & 2.2 & 0.7 & 0.6 & 0.4 & \\ 
 6 & 23.5 & 48.2 & 6.9 & 4.7 & 0.9 & 0.4 &  & \\ 
 7 & 26.1 & 44.1 & 8.6 & 4.3 & 1.1 & 0.2 & 0.6 & \\ 
 8 & 24.2 & 55.9 & 6.9 & 3.2 & 1.5 & 0.2 & 0.2 & \\ 
 9 & 26.6 & 49.7 & 8.0 & 3.9 & 1.5 & 0.4 &  & 0.2\\ 
 10 & 25.5 & 54.6 & 9.3 & 3.7 & 0.7 & 0.4 & 0.2 & \\ 
 11 & 29.2 & 57.5 & 7.1 & 3.9 & 1.7 & 0.4 & 0.2 & \\ \hline
total & 28.03 & 57.02 & 8.72 & 4.04 & 1.60 & 0.38 & 0.16 & 0.05\\ \hline
\end{tabular}

\caption{Degree distributions of loops as percentages. \label{pico_deg_perc}}
\end{table}

\begin{table}[!ht]
\tabletext
\begin{tabular}{|c|*{14}{c}c|} \hline
Index & 1 &  2 &  3 &  4 &  5 &  6 &  7 &  8 &  9 &  10 &  11 &  12 &  13 &  14 &  15 \\ \hline
 1  & 68 & 65 & 45 & 29 & 9 & 9 & 9 & 3 & 6 & 6 & 8 & 4 & 3 & 7 & 2 \\ 
 2  & 68 & 60 & 33 & 26 & 15 & 11 & 8 & 5 & 9 & 11 & 3 & 1 & 4 & 4 & 1 \\ 
 3  & 83 & 66 & 52 & 32 & 20 & 8 & 9 & 5 & 9 & 5 & 5 & 5 & 4 & 7 & 2 \\ 
 4  & 99 & 90 & 48 & 26 & 16 & 13 & 9 & 7 & 6 & 6 & 2 & 3 & 3 & 3 & 2 \\ 
 5  & 51 & 90 & 34 & 23 & 10 & 12 & 6 & 5 & 7 & 7 & 8 & 3 & 2 & 4 & 1 \\ 
 6  & 67 & 59 & 33 & 19 & 12 & 15 & 4 & 3 & 6 & 3 & 4 & 5 & 6 & 1 & 6 \\ 
 7  & 49 & 66 & 21 & 30 & 12 & 8 & 7 & 3 & 5 & 2 & 3 & 2 & 2 & 2 & 4 \\ 
 8  & 63 & 68 & 38 & 25 & 29 & 14 & 6 & 10 & 10 & 10 & 3 & 2 & 4 & 1 & 5 \\ 
 9  & 55 & 69 & 36 & 20 & 14 & 7 & 12 & 5 & 12 & 4 & 2 & 4 & 3 & 2 & 4 \\ 
 10  & 68 & 65 & 34 & 35 & 15 & 9 & 10 & 9 & 13 & 8 & 4 & 5 & 3 & 4 & 1 \\ 
 11  & 79 & 66 & 42 & 32 & 20 & 8 & 10 & 9 & 6 & 6 & 5 & 4 & 4 & 2 & 1 \\ 
\hline
\end{tabular}

\caption{Distribution of internal loops with at most 15 unpaired bases. \label{pico_regint}}
\end{table}

\begin{table}[!ht]
\tabletext
\begin{tabular}{|c|*{14}{c}c|} \hline
Index & 16 &  17 &  18 &  19 &  20 &  21 &  22 &  23 &  24 &  25 &  26 &  27 &  28 &  29 &  30 \\ \hline
 1  & & & 1 & & & 4 & 3 & & & & & & & & \\ 
 2  & 2 & 2 & 2 & 3 & 1 & 1 & 1 & 1 & 1 & 1 & & 1 & & & \\ 
 3  & 3 & 1 & 1 & & 1 & & & 1 & & & & & & 1 & \\ 
 4  & 3 & & & & & & & & 1 & & & & & & \\ 
 5  & & & & 1 & & & 2 & & & & & & 1 & & \\ 
 6  & & 2 & 1 & 3 & 3 & 1 & 2 & 1 & & & 1 & & & 1 & 1 \\ 
 7  & 3 & 1 & 5 & 2 & 2 & & 1 & 3 & 1 & & 1 & 2 & & & \\ 
 8  & 1 & 4 & 1 & 1 & 1 & & & 2 & 1 & & & & & & 1 \\ 
 9  & 5 & 3 & 5 & 2 & 1 & & 1 & 1 & & & & & & & \\ 
 10  & 3 & 2 & 1 & & 1 & & & & 1 & & & & 1 & & 1 \\ 
 11  & 2 & 3 & & 2 & 3 & & 1 & & & 1 & & & & 2 & 1 \\ 
\hline
\end{tabular}
 
\caption{Distribution of large ( $\geq$ 15 unpaired bases) internal loop sizes. \label{pico_exint}} 
\end{table}

\begin{table}[!ht]
\tabletext
\begin{tabular}{|c|c|*{7}{c}c|} \hline
Index & sum & 0 & 1 & 2 & 3 & 4 & 5 & 6 & 7\\ \hline
 1 & 366 & 142 & 148 & 41 & 20 & 12 & 3 &  & \\ 
 2 & 350 & 130 & 147 & 42 & 23 & 7 & 1 &  & \\ 
 3 & 381 & 138 & 171 & 44 & 16 & 11 & 1 &  & \\ 
 4 & 405 & 167 & 148 & 45 & 25 & 14 & 3 & 1 & 2\\ 
 5 & 341 & 136 & 126 & 58 & 12 & 4 & 3 & 2 & \\ 
 6 & 328 & 126 & 133 & 37 & 25 & 5 & 2 &  & \\ 
 7 & 341 & 140 & 122 & 46 & 23 & 6 & 1 & 3 & \\ 
 8 & 363 & 130 & 169 & 37 & 17 & 8 & 1 & 1 & \\ 
 9 & 361 & 143 & 143 & 43 & 21 & 8 & 2 &  & 1\\ 
 10 & 374 & 137 & 160 & 50 & 20 & 4 & 2 & 1 & \\ 
 11 & 392 & 157 & 164 & 38 & 21 & 9 & 2 & 1 & \\ \hline
total & 4002 & 1546 & 1631 & 481 & 223 & 88 & 21 & 9 & 3\\ \hline
\end{tabular}

\caption{Degree distributions of loops with contiguous stems. \label{pico_less3}}
\end{table}

\begin{table}[!ht]
\tabletext
\begin{tabular}{|c|*{7}{c}c|} \hline
Index & 0 & 1 & 2 & 3 & 4 & 5 & 6 & 7\\ \hline
 1 & 35.1 & 36.6 & 10.1 & 5.0 & 3.0 & 0.7 &  & \\ 
 2 & 31.8 & 35.9 & 10.3 & 5.6 & 1.7 & 0.2 &  & \\ 
 3 & 35.6 & 44.1 & 11.3 & 4.1 & 2.8 & 0.3 &  & \\ 
 4 & 48.0 & 42.5 & 12.9 & 7.2 & 4.0 & 0.9 & 0.3 & 0.6\\ 
 5 & 34.3 & 31.8 & 14.6 & 3.0 & 1.0 & 0.8 & 0.5 & \\ 
 6 & 30.7 & 32.4 & 9.0 & 6.1 & 1.2 & 0.5 &  & \\ 
 7 & 33.2 & 28.9 & 10.9 & 5.5 & 1.4 & 0.2 & 0.7 & \\ 
 8 & 32.0 & 41.6 & 9.1 & 4.2 & 2.0 & 0.2 & 0.2 & \\ 
 9 & 34.6 & 34.6 & 10.4 & 5.1 & 1.9 & 0.5 &  & 0.2\\ 
 10 & 33.9 & 39.6 & 12.4 & 5.0 & 1.0 & 0.5 & 0.2 & \\ 
 11 & 40.1 & 41.8 & 9.7 & 5.4 & 2.3 & 0.5 & 0.3 & \\ \hline
total & 38.63 & 40.75 & 12.02 & 5.57 & 2.20 & 0.52 & 0.22 & 0.07\\ \hline
\end{tabular}

\caption{Degree distributions of loops as percentages with contiguous stems. \label{pico_less3_perc}}
\end{table}

\clearpage

\pagebreak

\bibliographystyle{abbrv}
\bibliography{citeldp,temp,LDP}

\begin{thebibliography}{10}

\bibitem{yb-ceh-2}
Y.~Bakhtin and C.~E. Heitsch.
\newblock Large deviations for random trees.
\newblock Submitted.

\bibitem{cannone-etal-02}
J.~J. Cannone, S.~Subramanian, M.~N. Schnare, J.~R. Collett, L.~M. D'Souza,
  Y.~Du, B.~Feng, N.~Lin, L.~V. Madabusi, K.~M. M\"uller, N.~Pande, Z.~Shang,
  N.~Yu, and R.~R. Gutell.
\newblock The {C}omparative {RNA} {W}eb {(CRW)} {S}ite: an online database of
  comparative sequence and structure information for ribosomal, intron, and
  other {RNA}s.
\newblock {\em BMC Bioinformatics}, 3(1), 2002.

\bibitem{clote-etal-05a}
P.~Clote, L.~Gasieniec, R.~Kolpakov, E.~Kranakis, and D.~Krizanc.
\newblock On realizing shapes in the theory of {RNA} neutral networks.
\newblock {\em J. Theoret. Biol.}, 236(2):216--227, 2005.

\bibitem{clote-etal-06}
P.~Clote, E.~Kranakis, D.~Krizanc, and L.~Stacho.
\newblock Asymptotic expected number of base pairs in optimal secondary
  structure for random {RNA} using the {N}ussinov--{J}acobson energy model.
\newblock {\em Discrete Appl. Math}, 155(6-7):759--787, April 2007.

\bibitem{Dembo-Zeitouni:MR1619036}
A.~Dembo and O.~Zeitouni.
\newblock {\em Large deviations techniques and applications}, volume~38 of {\em
  Applications of Mathematics (New York)}.
\newblock Springer-Verlag, New York, second edition, 1998.

\bibitem{doshi-etal-04}
K.~J. Doshi, J.~J. Cannone, C.~W. Cobaugh, and R.~R. Gutell.
\newblock Evaluation of the suitability of free-energy minimization using
  nearest-neighbor energy parameters for {RNA} secondary structure prediction.
\newblock {\em BMC Bioinformatics}, 5(105), Aug 5 2004.

\bibitem{Ellis:MR2189669}
R.~S. Ellis.
\newblock {\em Entropy, large deviations, and statistical mechanics}.
\newblock Classics in Mathematics. Springer-Verlag, Berlin, 2006.
\newblock Reprint of the 1985 original.

\bibitem{fields-gutell-96}
D.~S. Fields and R.~R. Gutell.
\newblock An analysis of large {rRNA} sequences folded by a thermodynamic
  method.
\newblock {\em Fold Des}, 1(6):419--30, 1996.

\bibitem{fontana-etal-93b}
W.~Fontana, D.~Konings, P.~F. Stadler, and P.~Schuster.
\newblock Statistics of {RNA} secondary structures.
\newblock {\em Biopolymers}, 33(9):1389--404, Sept 1993.

\bibitem{insights}
C.~E. Heitsch.
\newblock Combinatorial insights into {RNA} secondary structures.
\newblock In preparation.

\bibitem{plane}
C.~E. Heitsch.
\newblock Combinatorics on plane trees, motivated by {RNA} secondary structure
  configurations.
\newblock Submitted.

\bibitem{hofacker-schuster-stadler-98}
I.~L. Hofacker, P.~Schuster, and P.~F. Stadler.
\newblock Combinatorics of {RNA} secondary structures.
\newblock {\em Discrete Appl. Math.}, 88(1-3):207--237, 1998.

\bibitem{lu-turner-mathews-06}
Z.~J. Lu, D.~H. Turner, and D.~H. Mathews.
\newblock A set of nearest neighbor parameters for predicting the enthalpy
  change of {RNA} secondary structure formation.
\newblock {\em Nucleic Acids Res}, 34(17):4912--4924, 2006.

\bibitem{mathews-etal-99}
D.~H. Mathews, J.~Sabina, M.~Zuker, and D.~H. Turner.
\newblock Expanded sequence dependence of thermodynamic parameters improves
  prediction of {RNA} secondary structure.
\newblock {\em J Mol Biol}, 288(5):911--940, May 21 1999.

\bibitem{miklos-meyer-nagy-04}
I.~Mikl{\'o}s, I.~M. Meyer, and B.~Nagy.
\newblock Moments of the {B}oltzmann distribution for {RNA} secondary
  structures.
\newblock {\em Bull. Math. Biol.}, 67(5):1031--1047, 2005.

\bibitem{nebel-04a}
M.~E. Nebel.
\newblock Identifying good predictions of {RNA} secondary structure.
\newblock In {\em Pac Symp Biocomput}, pages 423--34, 2004.

\bibitem{nebel-04b}
M.~E. Nebel.
\newblock Investigation of the {B}ernoulli model for {RNA} secondary
  structures.
\newblock {\em Bull. Math. Biol.}, 66(5):925--964, 2004.

\bibitem{palmenberg-sgro-97}
A.~C. Palmenberg and J.-Y. Sgro.
\newblock Topological organization of picornaviral genomes: Statistical
  prediction of {RNA} structural signals.
\newblock {\em S. Virol.}, 8:231--241, 1997.

\bibitem{rivas-eddy-00}
E.~Rivas and S.~R. Eddy.
\newblock Secondary structure alone is generally not statistically significant
  for the detection of noncoding {RNAs}.
\newblock {\em Bioinformatics}, 16(7):583--605, Jul 2000.

\bibitem{stanley-99}
R.~P. Stanley.
\newblock {\em Enumerative combinatorics. {V}ol. 2}, volume~62 of {\em
  Cambridge Studies in Advanced Mathematics}.
\newblock Cambridge University Press, Cambridge, 1999.
\newblock With a foreword by Gian-Carlo Rota and appendix 1 by Sergey Fomin.

\bibitem{tinoco-bustamante-99}
I.~{Tinoco, Jr} and C.~Bustamante.
\newblock How {RNA} folds.
\newblock {\em J Mol Biol}, 293(2):271--281, October 22 1999.

\bibitem{workman-krogh-99}
C.~Workman and A.~Krogh.
\newblock No evidence that {mRNAs} have lower folding free energies than random
  sequences with the same dinucleotide distribution.
\newblock {\em Nucleic Acids Res}, 27(24):4816--22, Dec 15 1999.

\bibitem{wuchty-etal-99}
S.~Wuchty, W.~Fontana, I.~L. Hofacker, and P.~Schuster.
\newblock Complete suboptimal folding of {RNA} and the stability of secondary
  structures.
\newblock {\em Biopolymers}, 49(2):145--65, Feb 1999.

\bibitem{zuker-03}
M.~Zuker.
\newblock Mfold web server for nucleic acid folding and hybridization
  prediction.
\newblock {\em Nucleic Acids Res}, 31(13):3406--15, 2003.

\bibitem{zuker-jaeger-turner-91}
M.~Zuker, J.~A. Jaeger, and D.~H. Turner.
\newblock A comparison of optimal and suboptimal {RNA} secondary structures
  predicted by free energy minimization with structures determined by
  phylogenetic comparison.
\newblock {\em Nucleic Acids Res}, 19(10):2707--2714, May 25 1991.

\bibitem{zuker-mathews-turner-99}
M.~Zuker, D.~Mathews, and D.~Turner.
\newblock Algorithms and thermodynamics for {RNA} secondary structure
  prediction: A practical guide.
\newblock In J.~Barciszewski and B.~Clark, editors, {\em RNA Biochemistry and
  Biotechnology}, NATO ASI Series, pages 11--43. Kluwer Academic Publishers,
  1999.

\end{thebibliography}

\end{document}